\documentstyle[12pt]{article}
\begin{document}

\title{New Non-Diagonal Singularity-Free Cosmological Perfect-Fluid Solution}

\author{Marc Mars  \thanks{Also at Laboratori de F\'{\i}sica Matem\`atica, 
IEC, Barcelona.} \\
Departament de F\'{\i}sica Fonamental, Universitat de Barcelona, \\
Diagonal 647, 08028 Barcelona, Spain.}
\date{}
\maketitle
\begin{abstract}
We present a new non-diagonal $G_2$ inhomogeneous perfect-fluid solution with
barotropic 
equation of state $p=\rho$ and positive density everywhere. It satisfies
the global hyperbolicity condition and has no curvature singularity
anywhere. This solution is very simple in form and has two arbitrary constants.

\end{abstract}

PACS Numbers: 04.20-Jb; 04.20-Cv; 98.80-Dr.

\vspace{2cm}

The study of exact inhomogeneous cosmological models is mainly motivated by the
fact that our Universe is not strictly spatially homogeneous. It was extensively
believed until very recently that all the cosmological models (homogeneous or
not) had to originate in a big-bang singularity whenever they satisfied an
energy condition for the matter contents and the model contained matter
everywhere. This belief was greatly reinforced by the powerful singularity
theorems {\cite {HP}}. However, in a remarkable letter published by J.M.M.
Senovilla in 1990 {\cite {S1}}, a new perfect-fluid inhomogeneous
cosmological solution without big-bang singularity and without
any other curvature singularity was presented. The matter
contents of that solution satisfied a realistic equation of state for radiation
dominated era $p=\rho/3$ with density (and therefore pressure) positive
everywhere.

In a subsequent paper {\cite{Sa}} this solution was shown to be
geodesically complete and satisfying causality conditions such as global
hyperbolicity. It was also proven that the singularity theorems for the existence of
incomplete geodesics could not be applied in that case because they need some
boundary or initial conditions, such as the existence of a causally trapped set,
which did not hold in that case. In consequence, the existence of physically
well-behaved cosmological solutions regular everywhere was clearly stated without
contradicting the powerful and aesthetically beautiful singularity theorems.

The above-mentioned singularity-free solution was generalized
in a paper {\cite {RS}} where the case of $G_2$ diagonal cosmologies assuming
separation of variables in co-moving coordinates was extensively studied. Among the
huge variety of perfect-fluid solutions with two commuting and mutually orthogonal
Killing vectors satisfying an Ansatz of separation of variables in a coordinate
system adapted to the fluid velocity vector, all the
different singular behaviours
were possible; there were solutions with a big-bang and/or big-crunch
singularity, solutions with timelike singularities but free of initial or final
singularities, solutions with pressure and density regular everywhere but with
singularities in the Weyl tensor, and finally a family of solutions completely
free of singularities of any type which contained the $p= \rho/3$ as a
particular case. It was also shown that this family was the most general
solution without singularities under the stated assumptions.

There have been some very recent attempts to prove that this family of
singularity-free solutions is unique in some sense. In fact, N. Dahdich
\& L.\,K. Patel {\cite {DP}} may have
proven that for diagonal cosmologies
\begin{eqnarray*}
ds^2= -A_0^2 dt^2 + A_1^2 dx_1^2 + A_2^2 dx_2^2 + A_3^2 dx_3^2
\end{eqnarray*}
with fluid velocity vector $\vec{u}$ parallel to $\partial_t$ (adapted
coordinates) and each of the functions $A_\alpha$ being a product of a function
of only $t$ with
a function of only the spatial coordinates (separation of variables) but without
assuming any symmetries in the space-time, the only singulariry-free
solution is given by the family pressented in {\cite {RS}} with a
two-dimensional group of isometries.

The purpose of this letter is to present a new perfect-fluid cosmological solution of
Einstein's equations without big-bang singularity or any other curvature singularities,
neither in the pressure or density nor in the Weyl tensor, satisfying a physically
well-motivated equation of state for a stiff fluid $p= \rho$, with density positive
and non-vanishing everywhere and satisfying a causality condition, namely global
hyperbolicity. This solution is very simple in form, and possesses a two-dimensional
abelian group of isometries acting on spacelike surfaces, but with neither of the
Killing vectors being hypersurface-orthogonal. Therefore, the metric is non-diagonal
and it belongs
to the class B(i) in Wainwright's classification of $G_2$ cosmologies \cite{W1}. 

The line-element of this solution is given by
\begin{eqnarray*}
ds^2= e^{\alpha a^2 r^2} \cosh(2at) \left ( -dt^2 + dr^2 \right ) + r^2 \cosh(2at)
d\phi^2 + \frac{1}{\cosh(2at)} \left ( dz + a r^2 d\phi \right )^2,
\end{eqnarray*}
where $\alpha$ and $a$ are arbitrary constants, and the range of variation of the
coordinates is
\begin{eqnarray*}
-\infty < t,z < \infty \hspace{1cm} & 0\leq r < \infty \hspace{1cm} & 0 \leq
\phi \leq 2 \pi.
\end{eqnarray*}
This spacetime has a well-defined axis of symmetry at $r=0$ where the
so-called elementary flatness {\cite{KK}} is safisfied and therefore the
coordinate $r$ has to be interpreted as a radial cylindrical coordinate. The
energy-momentum tensor of this spacetime corresponds to a perfect fluid
with velocity vector given by 
\begin{eqnarray*}
\vec{u} = \frac{e^{-\frac{1}{2} \alpha a^2 r^2}}{\cosh^{1/2}(2at)}
 \partial_t
\end{eqnarray*}
and density and pressure
\begin{eqnarray*}
p=\rho = \frac{a^2 \left ( \alpha -1 \right ) e^{-\alpha a^2 r^2}}{\cosh(2at)},
\end{eqnarray*}
which are positive everywhere whenever the constant $\alpha$ is restricted to
\begin{eqnarray*}
\alpha > 1.
\end{eqnarray*}
It can be seen that this $p=\rho$ solution can be generated from a vacuum
solution using the generating procedure of Wainwright, Ince \& Marshman
{\cite {WW}}. The vacuum
solution that leads to this spacetime is contained in this family as the
special case $\alpha=1$.

It is trivial to check that (when the constant $a$ is not zero) the only Killing
vectors of this metric are given by  $\partial_z$ and $\partial_\phi$, except in 
the particular case $\alpha=0$ in which the metric has a four-dimensional
group of symmetries acting on three-dimensional spacelike hypersurfaces and that
contains a three-dimensional subgroup belonging to Bianchi type II. Thus,
in this particular case the solution is locally rotationally symmetric.
 
In the natural null tetrad that can be read from the expression
of the line-element, the non-vanishing components of the Weyl tensor are given by
\begin{eqnarray*}
\Psi_0  &=&\frac{a^2 e^{-\alpha a^2 r^2}}{\cosh^3(2at)} \left [
 \cosh^2(2at)(\alpha/2 + 1) + \alpha a r \cosh(2at) \sinh(2at) - 3 - \right. \\
& & \hspace{2cm} \left . i (\alpha a r \cosh(2at)+ 3 \sinh(2at) ) \right ]  \\
\Psi_2 &=&  \frac{a^2 e^{-\alpha a^2 r^2}}{\cosh^3(2at)} \left [
\cosh^2(2at)(\alpha/6 + 1/3) - 1 - i \sinh(2at)  \right ] \\ 
\Psi_4 &=& \frac{a^2 e^{-\alpha a^2 r^2}}{\cosh^3(2at)} \left [
 \cosh^2(2at) (\alpha/2 + 1)- \alpha a r \cosh(2at) \sinh(2at) - 3 - \right. \\
& & \hspace{2cm} \left . i (-\alpha a r \cosh(2at)+ 3 \sinh(2at) ) \right ],
\end{eqnarray*}
which show that this solution has Petrov type I everywhere except on the axis
of symmetry where it degenerates to type D. From the expressions above for the
energy density, pressure and Weyl tensor, it is obvious that this spacetime
has no curvature singularities anywhere. In fact, an easy 
calculation shows
that not only all the scalar
polynomials in the Riemann tensor are regular, but all scalars based on the 
derivatives of Riemann are as well. 

In order to see that a causality condition is satisfied, it is
sufficient to note that the coordinate $t$ in the metric is a time function, that
is, it increases along every future-directed non-spacelike curve.  This can
be checked by computing the gradient of this function which results timelike
everywhere. The existence of a time function is the necessary and sufficient
condition for the stable causality condition to hold. Using a result due to
R. Geroch (\cite {Ge}) concerning the null geodesics and the fact that this
metric is causally stable, it can be easily proven that this spacetime
is in fact globally hyperbolic. A detailed study of the geodesics
of this spacetime shows that the solution is geodesically complete and therefore
singularity-free.

We will use the natural orthonormal tetrad of the line-element 
to write the components of the kinematical tensors associated with the
fluid velocity vector $\vec{u}$.
The expansion and the non-zero components of the shear tensor
of the fluid velocity vector are given by
\begin{eqnarray*}
\theta = \frac{ a e^{-\frac{1}{2}\alpha a^2 r^2 } \sinh(2at)}{\cosh^{3/2}(2at)},
\hspace{15mm}
\sigma_{11}= \sigma_{22}= -\frac{\sigma_{33}}{2} = \frac{2}{3} \theta.
\end{eqnarray*}

The vorticity of $\vec{u}$ vanishes obviously and its acceleration is
\begin{eqnarray*}
\vec{a} = \alpha a^2 r\frac{e^{-\alpha a^2 r^2}}{\cosh(2at)} \partial_r.
\end{eqnarray*}

From the expressions of the pressure, density, Weyl tensor and kinematical
quantities, it follows that
this solution is nearly flat when $t \rightarrow - \infty$, where all these
quantities tend to zero. Then, as $t$ increases, the fluid begins to contract
while the density obviously increases. This contraction lasts until $t=0$ when
the fluid begins  to expand. At this time the density reaches a maximum along
the world lines of the fluid and then starts to decrease. This maximum can be
made arbitrarily large by fixing the constant $\alpha$ big enough. As $t
\rightarrow \infty $ the solution tends again to a nearly flat situation. The
avoidance of collapse into a singularity during the contracting era is
provided by the radial gradient of pressure which produces an eternal
outwards acceleration onthe fluid. Thus, the inhomogeneity of the spacetime
allows it to avoid any collapse into a singularity.

The deceleration parameter can be computed from the expression for the
expansion and gives
\begin{eqnarray*}
q= 8 - 6 \frac{\cosh^2(2at)}{\sinh^2(2at)},
\end{eqnarray*}
which shows that this solution has an inflationary epoch near the rebound time
$t=0$ while it is non-inflationary for the rest of its history.

Now, we will write the diagonal limit of this solution. This limit
is obtained when the constant $a$, which essentially gives a scale to the
spacetime, is zero. This limit can be made maintaining the product
$ \alpha a^2 = \beta $ constant, thus leading to the metric
\begin{eqnarray*}
ds^2= e^{\beta r^2} \left ( -dt^2 + dr^2 \right ) + r^2 d\phi^2
+  dz^2.
\end{eqnarray*}
This metric is obviously static and its matter contents is a stiff
perfect-fluid with pressure and density given by
\begin{eqnarray*}
p=\rho= \beta e^{-\beta r^2}.
\end{eqnarray*}
The non-zero components of the Weyl tensor in this particular case are
\begin{eqnarray*}
\Psi_0 = \Psi_4 = 3 \Psi_2 = \frac{1}{2} \beta e^{-\beta r^2},
\end{eqnarray*}
so that this metric is type D everywhere. In consequence, it belongs to
a class due to Barnes {\cite{B1}}. Being static and cylindrically symmetric
it also belongs to the general solution found by K.A. Bronnikov (\cite{BB})
and rediscovered by D. Kramer (\cite{K1}).
Finally, the vacuum limit of this solution ($\beta=0$) is the Minkowski
spacetime.

Summing up, we have found the first non-diagonal singularity-free cosmological
model satisfying physically well-behaved properties, and besides, we have shown
that the existence of singularity-free models is likely to occur
when more general than separable $G_2$ diagonal cosmologies are considered.
This solution has been found by studying the perfect-fluid orthogonally
transitive $G_2$ cosmologies with no hypersurface-orthogonal Killing vector 
( class B(i) in Wainwright's classification) with an Ansatz of
separation variables in co-moving coordinates. This case has been in fact
exhausted and the results will be published in the near future.

\section*{Acknowledgements}
The author wishes to thank the {\it Direcci\'o General d'Universitats, Generalitat
de Catalunya}, for financial support.

All the tensors in this paper have been computed with the
algebraic computer programs CLASSI and REDUCE.

\end{document}